\documentclass[fleqn,10pt]{wlscirep}
\usepackage[utf8]{inputenc}
\usepackage[T1]{fontenc}

\usepackage{graphicx}
\usepackage[most]{tcolorbox}
\usepackage{enumitem}
\usepackage{xcolor}
\usepackage{caption}
\usepackage{subcaption}
\usepackage{amsmath} 
\usepackage{dcolumn}
\usepackage{bm}

\usepackage{multirow}

\title{Non-intrusive Monitoring of Sealed Microreactor Cores Using Physics-Informed Muon Scattering Tomography With Momentum Measurements}

\author[1]{Reshma Ughade}
\author[1,*]{Stylianos Chatzidakis}
\affil[1]{School of Nuclear Engineering, Purdue University, West Lafayette, IN  47907, USA}
\affil[*]{schatzid@purdue.edu}

\begin{abstract}

Next-generation advanced reactors, including microreactors, enable remote deployment and semi-autonomous operation with reduced staffing. Their compact, sealed, and highly heterogeneous cores challenge traditional safeguards and monitoring approaches developed for large reactors, spent-fuel storage, and fuel-cycle facilities, where verification relies on access to declared materials, containment and surveillance, or bulk accountancy methods. These approaches do not readily translate to microreactors, where sealed cores, transportability, limited inspection access, and complex internal geometries constrain conventional verification strategies and reduce sensitivity to localized anomalies such as missing fuel. In this work, we demonstrate muon-based missing-fuel detection in microreactor-scale geometries under realistic cosmic-ray conditions. We introduce $\mu$TRec, a physics-informed muon scattering tomography framework that reconstructs event-level curved muon trajectories using a Gaussian multiple Coulomb scattering model with Bayesian updating and maps scattering density through voxel-wise M-values to enable core integrity verification. The method is evaluated on a representative hexagonal microreactor core containing 61 fuel flakes with embedded control drums and shutdown rods, using both idealized 5 GeV beams and realistic 0--60 GeV zenith-angle-dependent cosmic-ray muon spectra. A missing fuel flake is detected using $3\times10^6$ muons at 50 mm voxel resolution. Incorporating per-muon momentum increases detectability by up to 149.85\% for the laser-driven source and 105.11\% for the cosmic-ray source, relative to momentum-agnostic reconstruction, while preserving structural visibility across voxel sizes. The framework is robust to practical detector constraints, with only an 8.88\% reduction in detectability for 10 mm spatial resolution and 10\% energy resolution. Compared with conventional PoCA reconstruction, $\mu$TRec achieves 326.13\% to 392.14\% higher detectability in the momentum-informed case at equivalent muon counts, enabling defect identification at reduced acquisition times.

\end{abstract}
\begin{document}

\flushbottom
\maketitle

\section*{Introduction}

Microreactors are an emerging class of nuclear energy systems that could expand the range of applications for clean, resilient power by enabling deployment at much smaller scales than traditional plants \cite{osti_1806274}. Microreactors are commonly categorized by their electrical output, typically in the range of 1--20 MWe \cite{TESTONI2021103822} and a defining attribute is portability. Many designs are intended to be factory fabricated and transported to the point of use, allowing siting in locations where large commercial reactors are impractical or prohibitively expensive. Consequently, microreactors are being considered for long duration electricity and heat in remote and energy constrained settings, including diesel dependent communities (e.g., Arctic towns and island grids), critical facilities where outages carry high consequences (hospitals, emergency operations centers, and water treatment plants), and disaster recovery contexts requiring sustained replacement power after major disruptions \cite{Black31012023, LANE2025105520, MORONESGARCIA2024113021, Stevens01122024}. Beyond terrestrial applications, their compact and potentially autonomous operation has been proposed for off-grid power in extreme environments, such as surface power systems for future space missions \cite{Watson18042025}.

Because microreactors are designed to be transportable, sealed, and capable of operation with reduced on-site staffing, they introduce new challenges for safeguards and monitoring \cite{Ryan31072024, Culler2024RiskAnalysisRemoteMicroreactors}. Existing safeguards frameworks were primarily developed for large power reactors, spent-fuel storage, and fuel-cycle facilities, where verification relies on access to declared materials, operator records, containment and surveillance systems, and bulk accountancy measurements. These approaches are effective for tracking material flows and verifying facility-level inventories but are less suited to compact reactor systems with sealed cores and limited inspection access. Microreactors may be deployed in remote locations, relocated over their lifetime, or operated with minimal on-site infrastructure, constraining opportunities for frequent inspection and direct measurements. In addition, their highly heterogeneous internal configurations, including heat pipes, control drums, reflectors, and distributed fuel elements, reduce the sensitivity of bulk verification approaches to localized anomalies. As a result, detecting configuration-level changes, such as missing fuel or unauthorized modifications within a sealed core, remains a significant safeguards challenge \cite{Bryan2023RemoteMicroreactors}.

These constraints motivate the development of non-intrusive monitoring techniques capable of interrogating the internal configuration of sealed microreactor cores without reliance on direct access or operator declarations. Muon scattering tomography is a promising candidate because it leverages naturally occurring cosmic-ray muons that penetrate thick shielding and dense reactor structures, enabling volumetric inference of internal material distributions \cite{Borozdin_2003, Miyadera:22, liu_muon_2018, chatzidakis_interaction_2016, chatzidakis_analysis_2016}. This capability is particularly relevant for safeguards scenarios where the objective is verification of core integrity and detection of localized anomalies, such as fuel relocation, missing components, or unexpected high-Z additions, in systems that cannot be easily opened or instrumented. Cosmic-ray muons are highly penetrating charged particles produced in the atmosphere that reach the ground with a broad energy spectrum and can traverse meters of dense material \cite{bae_gamma-ray_2024}. Muon tomography exploits this penetrability by measuring the incoming and outgoing trajectories of individual muons and inferring the intervening material distribution from either attenuation (absorption-based imaging) or angular deflections due to multiple Coulomb scattering (MCS) \cite{PhysRevLett.109.152501}. Because scattering depends strongly on material properties and muon momentum, the technique provides volumetric sensitivity to changes in effective density and atomic number, even in shielded or inaccessible systems.

To enable safeguards-relevant monitoring of microreactor cores, we develop a physics-informed muon scattering tomography framework tailored to microreactor geometries and demonstrate muon-based detection of missing fuel under ideal and realistic cosmic-ray conditions. We leverage our prior work on $\mu$TRec \cite{ughade_trec_2025, ughade_performance_2023}, a physics-informed trajectory estimation and reconstruction method that explicitly models MCS rather than assuming straight-line transport. For each detected muon, $\mu$TRec uses the measured entry and exit track segments and, when available, event-level momentum information to estimate a most probable curved trajectory under a Gaussian scattering model with Bayesian updating and propagates the associated uncertainty into voxelized image reconstruction. The key innovation of $\mu$TRec is the treatment of the muon path as a probabilistic, physics-constrained trajectory rather than a single point or straight chord through the inspection region. Compared with classical methods such as point-of-closest-approach and other straight-line approximations, this approach reduces localization bias, improves reconstruction fidelity in heterogeneous geometries, and enhances sensitivity to small configuration changes from limited muon statistics. These capabilities are essential for safeguards applications, where the objective is not only visualization but reliable detection and localization of anomalies within sealed cores. By enabling detection of missing fuel and internal configuration changes under realistic cosmic-ray conditions, physics-informed muon tomography provides a pathway toward deployable, non-intrusive verification tools for next-generation microreactor systems \cite{ughade2023muon, Ughade_assessment, ughade_physics-based_2023, ughade_--fly_2023, bae_gamma-ray_2024, ughade_3d_2024, chatzidakis_generalized_2018, Ughade_efficient_2025, Ughade_resolution_2025}.

In this work, we extend physics-informed muon scattering tomography to microreactor-scale reactor systems and safeguards-driven monitoring scenarios. Unlike prior $\mu$-tomography studies focused on cargo screening, geological imaging, or spent-fuel verification, this work targets compact, sealed, and highly heterogeneous microreactor cores, where limited inspection access and transportability create distinct verification challenges. Second, we demonstrate detection of missing fuel in a representative microreactor geometry under both idealized beam conditions and realistic cosmic-ray flux. We show that configuration-level anomalies can be identified in sealed cores using achievable muon counts, providing a practical pathway for non-intrusive core integrity verification. Finally, we quantify performance under realistic detector and operational constraints, including limited spatial resolution, energy resolution, and muon statistics. Results show that the proposed approach maintains anomaly sensitivity and reduces acquisition time requirements relative to conventional, point of closest approach (PoCA)-based methods, supporting feasibility for field deployment.

In this paper, the Methodology section describes the microreactor geometry used in this study, the simulated muon source and its characteristics in GEANT4, and the detector specifications. It also briefly describes the PoCA algorithm and then presents the physics-informed $\mu$TRec framework, with emphasis on how momentum information is incorporated into the algorithm. The $M$-value and detector resolution definitions are also provided. In the Results and Discussion section, five scenarios are considered to evaluate the developed $\mu$TRec algorithm, including a laser-based muon source, a cosmic-ray muon source, a $\mu$TRec versus PoCA comparison, the effect of detector resolution, and the effect of the number of spectrometers.

\section*{Methodology}

\subsection*{eVinci$^{TM}$ heat pipe microreactor}

\begin{figure}[h]
\centering
\begin{subfigure}[b]{0.8\linewidth}
    \centering
    \includegraphics[width=\linewidth]{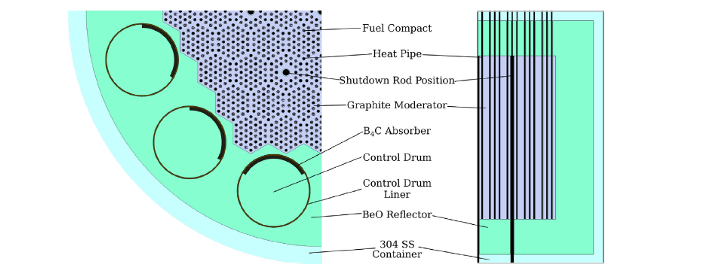}
    \caption{}
    \label{microreactor_geometry_a}
\end{subfigure}
\hfill
\begin{subfigure}[b]{0.6\linewidth}
    \centering
    \includegraphics[width=\linewidth]{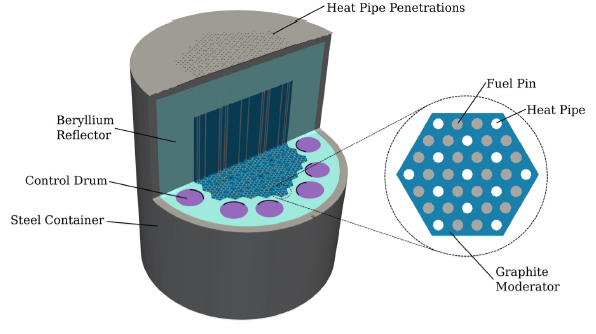}
    \caption{}
    \label{microreactor_geometry_b}
\end{subfigure}

\caption{EMD microreactor: (a) radial and axial diagram with labeled core components \cite{PRICE2023112709} and (b) basic layout with enlarged fuel flake \cite{PRICE2024134}. }
\label{fig:microreactor_geometry}
\end{figure}

To enable an industry relevant study while avoiding proprietary design details, we developed a heat-pipe microreactor proxy inspired by the eVinci concept under development at Westinghouse Electric Company \cite{10.1115/ICONE28-67519}. The resulting configuration is graphite moderated, fueled with UO$_2$ , and controlled using 12 rotating control drums. Throughout this work, the model is referred to as the eVinci-motivated design (EMD) microreactor. A labeled figure of the EMD microreactor core layout is provided in Figure \ref{microreactor_geometry_a}. The arrangement of the UO$_2$ fuel compacts and heat pipes within a representative fuel flake is shown in Figure \ref{microreactor_geometry_b}. Additional design details for the EMD are summarized in Table \ref{tab:emd_params}. The reference configuration includes no burnable absorbers.

\begin{table}[ht]
\centering
\begin{tabular}{|l|c|c|}
\hline
\textbf{Parameter} & \textbf{Value} & \textbf{Unit} \\
\hline
Fuel Pin Diameter & 1.7 & [cm] \\
\hline
Active Fuel Height & 182 & [cm] \\
\hline
Drum Liner Outer Radius & 18.7 & [cm] \\
\hline
Drum Liner Inner Radius & 18.05 & [cm] \\
\hline
Drum Liner Material & lead & NA \\
\hline
Drum B$_4$C Coating Angle & 110 & [$^\circ$] \\
\hline
Drum B$_4$C Coating Thickness & 1.5 & [cm] \\
\hline
Drum Height & 182 & [cm] \\
\hline
Heat Pipe Outer Diameter & 1.6 & [cm] \\
\hline
Heat Pipe Wall Thickness & 1.0 & [mm] \\
\hline
Heat Pipe Material & SS & NA \\
\hline
Shutdown Guide Tube Outer Radius & 2.0 & [cm] \\
\hline
Shutdown Guide Tube Thickness & 0.3 & [cm] \\
\hline
Shutdown Guide Tube Material & SS & NA \\
\hline
Pin Cell Pitch & 2.86 & [cm] \\
\hline
Flake Pitch & 18 & [cm] \\
\hline
Reflector Outer Radius & 120 & [cm] \\
\hline
Container Height & 280 & [cm] \\
\hline
Container Outer Radius & 130 & [cm] \\
\hline
Container Thickness & 10 & [cm] \\
\hline
\end{tabular}
\caption{\label{tab:emd_params}Design parameters for the EMD microreactor. All pitches are reported as the flat-to-flat distances for the hexagonal array \cite{PRICE2023112709}.}
\end{table}

\subsection*{Muon source}


Two distinct muon sources were modeled to evaluate performance under both controlled-beam and realistic background conditions. The first source represents a laser-generated muon beam, implemented as a parallel source incident on the inspection volume. Muon energies were sampled about an average energy of 5 GeV. This source provides a well-defined illumination geometry and reduced angular spread, enabling controlled studies of reconstruction sensitivity, acquisition requirements, and algorithm performance under near-ideal beam conditions.

\begin{figure}[h]
\centering
\begin{subfigure}[b]{0.45\linewidth}
    \centering
    \includegraphics[width=\linewidth]{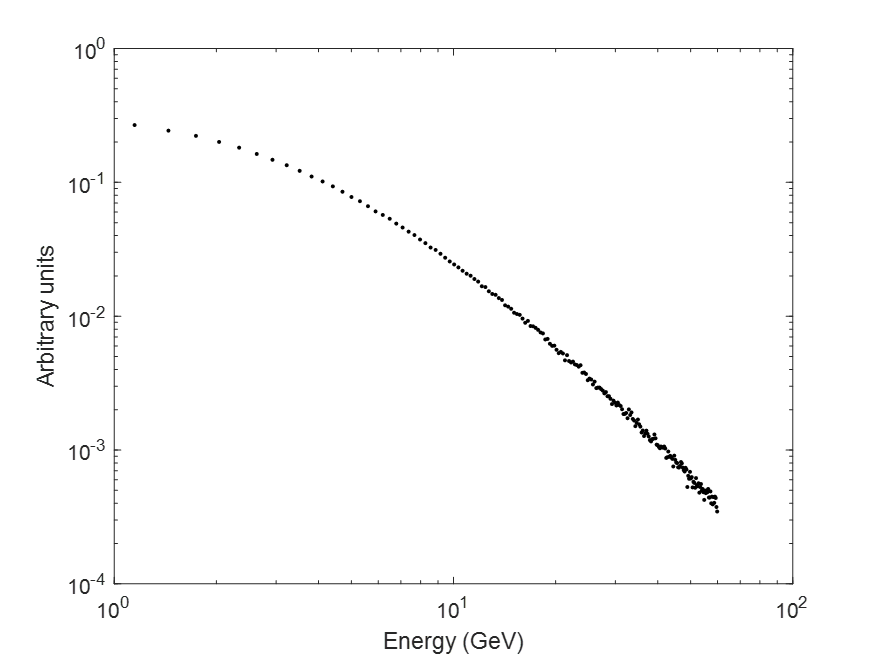}
    \caption{Energy distribution of simulated muon particles.}
\end{subfigure}
\hfill
\begin{subfigure}[b]{0.45\linewidth}
    \centering
    \includegraphics[width=\linewidth]{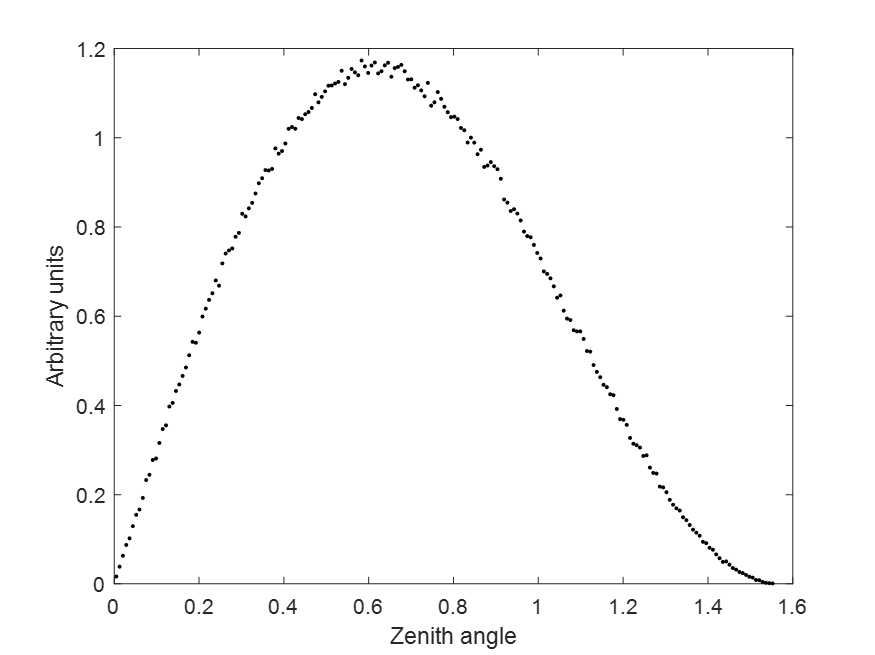}
    \caption{Angular distribution (radians) of simulated muon particles.}
\end{subfigure}

\caption{Simulated cosmic-ray muon source characteristics used in GEANT4: (a) energy and (b) angular (zenith) distributions \cite{chatzidakis_geant4-matlab_2015}.}
\label{fig:muon_source}
\end{figure}

The second source represents naturally occurring cosmic-ray muons and was simulated in GEANT4 using the Muon Generator described in Reference \cite{chatzidakis_geant4-matlab_2015}. For this configuration, muon energies span 0–60 GeV and the incident directions are sampled over a zenith-angle range of 0–90°, providing a broad spectrum in both energy and solid angle consistent with background muon fields. The resulting source distributions are summarized in Figure \ref{fig:muon_source}, which show the simulated energy spectrum and angular spread used for subsequent imaging and reconstruction studies.

\subsection*{Detectors}

Muon tracking in this study is performed using a four plane scintillation detector array. Each detector plane has dimensions of 300$\times$300$\times$1 cm$^3$ and is oriented vertically. The overall separation between the first and fourth planes is 400 cm, with the detectors grouped into upstream and downstream pairs. Within each pair, the inter-plane spacing is 30 cm, enabling estimation of the incoming and outgoing track segments used for trajectory reconstruction. For most results in this analysis, the detector planes are assumed to have ideal spatial and energy resolution, so the reconstruction performance can be evaluated independently of detector measurement uncertainty. To assess practical detector effects, we also examined two representative finite resolution cases by introducing uncertainty in the reconstructed muon hit positions and muon energy.

\begin{figure}[h]
\centering
\begin{subfigure}[b]{0.45\linewidth}
    \centering
    \includegraphics[width=\linewidth]{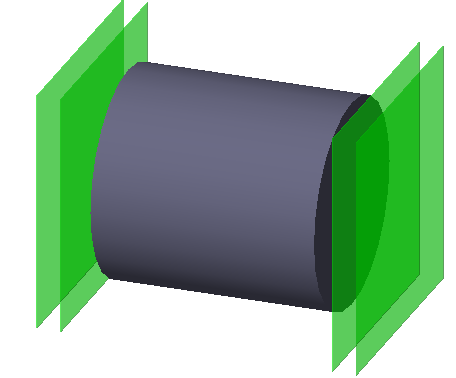}
    \caption{Geometry and detector setup.}
\end{subfigure}
\hfill
\begin{subfigure}[b]{0.35\linewidth}
    \centering
    \includegraphics[width=\linewidth]{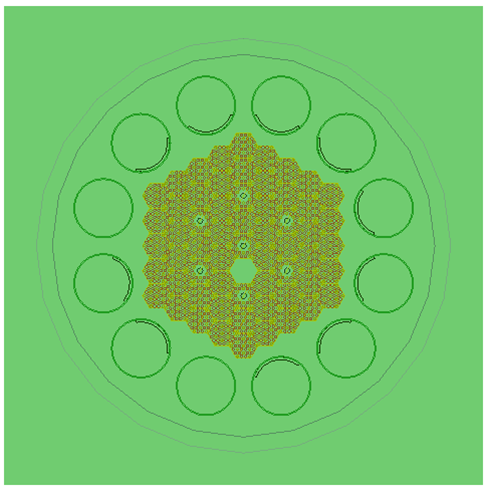}
    \caption{Side view of the geometry and detector setup.}
\end{subfigure}
\caption{Measurement setup of EMD microreactor (grey) modelled in GEANT4 for muon entry/exit tracking using four detectors (green).}
\label{fig:microreactor}
\end{figure}

All simulations are performed using GEANT4 simulation software \cite{agostinelli_geant4simulation_2003}. The overall imaging configuration is shown in Figure \ref{fig:microreactor}. Muons traverse the upstream detector pair, pass through the microreactor geometry where they undergo MCS governed by the materials encountered, and then traverse the downstream detector pair. The four detector planes provide the position measurements required to estimate the incoming and outgoing track segments used in reconstruction.

In addition to track geometry, we investigate the benefit of incorporating muon momentum information. In practical systems, momentum is commonly measured using dedicated spectrometers, for example Cherenkov or magnetic spectrometers, which can increase detector complexity and cost \cite{bae_fieldable_2022}. Because the simulation provides access to the muon’s initial and final energies, we emulate momentum measurement under two scenarios: (i) momentum available only at entry, and (ii) momentum available at both entry and exit, corresponding to a two-spectrometer configuration. In our implementation, the momentum measurements are associated with detector 2 (entry side) and detector 3 (exit side). The entry-only case supplies the per-muon momentum that can be incorporated in the $\mu$TRec algorithm and to compute the density mapping (M-value), while the entry-and-exit case additionally provides an estimate of energy loss through the object for inclusion in $\mu$TRec. As expected, increasing the available measurement information improves reconstruction fidelity and reduces ambiguity in the inferred material distribution.

\subsection*{Muon scattering tomography}
In this study, we employ the $\mu$TRec algorithm, which provides a physics-informed alternative to the purely geometric PoCA approach and has demonstrated substantially improved computational efficiency, up to approximately 20× faster in prior evaluations \cite{Ughade_efficient_2025, ughade_trec_2025}. To quantify performance, reconstructions obtained with $\mu$TRec are directly compared against PoCA results. 

\subsubsection*{PoCA algorithm}

To reconstruct the scattering locations of muons within an object, the PoCA algorithm adopts a geometrically simplified model \cite{SCHULTZ2004687}. This method operates under the assumption that the total angular deflection a muon experiences while traversing a medium can be attributed to a single discrete scattering event. Unlike more sophisticated models that attempt to capture the cumulative effects of MCS along the muon path, PoCA approximates the entire deviation by assigning a single point to represent the interaction. The implementation begins with the determination of incoming and outgoing muon trajectories based on position measurements collected from two pairs of tracking detectors. It is assumed that the detectors have sufficient resolution to accurately associate each recorded hit with its corresponding muon track. From this data, two vectors are constructed to represent the entry and exit paths. The intersection of these two vectors is considered the scattering point. However, these vectors may not intersect in three-dimensional space. Instead, the PoCA point is computed by identifying the pair of points, one on each trajectory, that are closest to one another in Euclidean distance. This is accomplished using a least-squares minimization of the perpendicular distance between the two lines. The midpoint of this shortest segment is then defined as the scattering point, and the scattering angle is estimated as the angular difference between the two trajectory vectors. Because PoCA is a purely geometric, simplified trajectory model, it does not provide a natural way to incorporate per-muon momentum into the reconstruction itself. In practice, momentum information can only be used for density mapping through the $M$-value. When momentum is unavailable, density mapping is performed using the measured scattering angles alone.

\subsubsection*{$\bm{\mu}$TRec algorithm}

To overcome the limitations of single point scattering approximations, the $\mu$TRec algorithm reconstructs the most probable muon trajectory by explicitly incorporating MCS physics. Rather than assigning a single interaction point, $\mu$TRec models the muon path through the interrogation volume as a probabilistic curve constrained by the measured upstream and downstream track segments. The scattering process is described through an MCS likelihood, in which the expected angular spread depends on the muon momentum and the material traversed. A commonly used approximation for the root mean square (RMS) scattering angle over a thickness $z$ is by the formula \cite{HIGHLAND1975497, LYNCH19916},
\begin{equation}
\sigma_\theta \approx \frac{13.6~\mathrm{MeV}}{\beta p c}\sqrt{\frac{z}{X_0}}
\left[1 + 0.038\ln\!\left(\frac{z}{X_0}\right)\right],
\end{equation}
where $p$ is the muon momentum, $\beta c$ is the muon speed, and $X_0$ is the radiation length of the material. This relation directly motivates the use of momentum information in trajectory inference, since higher momentum muons undergo less scattering and their measured deflections should therefore be weighted differently from those of lower momentum muons.

The $\mu$TRec method adopts a generalized statistical framework for cosmic ray muon tomography based on the formalism introduced in Refs.~\cite{chatzidakis_generalized_2018, schulte_maximum_2008, ughade_trec_2025, ughade_performance_2023}. In each transverse projection, MCS is represented using a bivariate Gaussian model that jointly describes angular deflection ($\theta$) and lateral displacement ($y$). For the $y$--$z$ projection, the state vector is written as
\begin{equation}
\mathbf{Y}(z)=
\begin{bmatrix}
y(z)\\
\theta_y(z)
\end{bmatrix}.
\end{equation}
An analogous definition is used in the $x$--$z$ projection.

Energy loss is incorporated using the continuous slowing down approximation (CSDA), which is appropriate for minimum ionizing muons over the energy range relevant to this work. In contrast to straight line or single point scattering models, $\mu$TRec estimates a continuous curved trajectory that more faithfully represents the physical interactions experienced by muons in dense and nonuniform media. The reconstructed trajectory asymptotically approaches the measured incoming and outgoing tracks, which improves spatial localization and image fidelity.

\begin{figure}[htbp]
    \centering
    \includegraphics[width=0.5\linewidth]{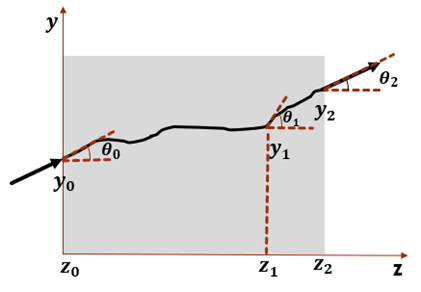}
    \caption{Illustration of multiple Coulomb scattering in the $y$--$z$ plane \cite{ughade_trec_2025}.}
    \label{muTRec_derivation}
\end{figure}

We consider the $y$--$z$ projection of the reconstruction volume, shown in Fig.~\ref{muTRec_derivation}, where $(y_0, z_0, \theta_0)$ denotes the muon state at the entry of the reconstruction volume and $(y_2, z_2, \theta_2)$ denotes the state at the exit. The $z$ direction is normal to the detector planes.

The $\mu$TRec estimate in the $y$--$z$ plane is given by \cite{ughade_trec_2025}
\begin{equation}\label{eq:y_muTRec}
\mathbf{Y}_{\mu\mathrm{TRec}}(z)=
\left( \Sigma_1^{-1} + R_2^{T}\Sigma_2^{-1}R_2 \right)^{-1}
\left( \Sigma_1^{-1}R_0\mathbf{Y}_0 + R_2^{T}\Sigma_2^{-1}\mathbf{Y}_2 \right),
\end{equation}
and similarly, in the $x$--$z$ plane,
\begin{equation}\label{eq:x_muTRec}
\mathbf{X}_{\mu\mathrm{TRec}}(z)=
\left( \Sigma_1^{-1} + R_2^{T}\Sigma_2^{-1}R_2 \right)^{-1}
\left( \Sigma_1^{-1}R_0\mathbf{X}_0 + R_2^{T}\Sigma_2^{-1}\mathbf{X}_2 \right).
\end{equation}
Here, $\Sigma_1$ and $\Sigma_2$ are covariance matrices for the upstream and downstream trajectory segments, respectively, and $R_0$ and $R_2$ are the corresponding transport matrices. Full definitions of these matrices, along with the derivation of Eqs.~\eqref{eq:y_muTRec} and \eqref{eq:x_muTRec}, are provided in Ref.~\cite{ughade_trec_2025}. A pseudocode summary of the momentum-informed $\mu$TRec workflow is shown in Fig.~\ref{fig:mutrec-pseudocode}.

\begin{figure}[htbp]
\centering
\begin{tcolorbox}[
  enhanced,
  width=0.93\linewidth,
  colback=gray!10,
  colframe=black,
  boxrule=0.8pt,
  arc=3mm,
  left=4mm,
  right=4mm,
  top=3mm,
  bottom=3mm
]
\small
\begin{enumerate}[label=\arabic*., leftmargin=*, itemsep=3pt, topsep=0pt, parsep=0pt, partopsep=0pt]

\item Collect incident and exiting muon measurements:
\begin{center}
$(p_1,p_2,E_1)$ from the upstream tracker,\\
$(p_3,p_4,E_2)$ from the downstream tracker.
\end{center}

\item Compute the scattering angle and energy loss $(E_1-E_2)$ for each muon.

\item Estimate the muon trajectory using a Gaussian MCS model and Bayesian formalism \cite{ughade_trec_2025}:
\begin{enumerate}[label=(\alph*), leftmargin=2.2em, itemsep=2pt, topsep=2pt, parsep=0pt]
    \item Initialize the radiation length estimate.
    \item Apply a linear energy loss model.
    \item Compute MCS moments and covariance terms.
    \item Construct covariance matrices $\Sigma_1(z)$ and $\Sigma_2(z)$, and transport matrices $R_1(z)$ and $R_2(z)$.
    \item Apply the Bayesian update to obtain the estimated trajectory.
\end{enumerate}

\item Discretize the reconstruction volume into $n \times n \times n$ voxels:
\begin{enumerate}[label=(\alph*), leftmargin=2.2em, itemsep=2pt, topsep=2pt, parsep=0pt]
    \item Assign an $M$-value (Eq.~\ref{Mvalue}) to each voxel intersected by the estimated trajectory.
    \item Repeat for all muons and average voxel statistics to form the $M$-value density map.
\end{enumerate}

\item Reconstruct the image from the voxelized statistics.

\end{enumerate}
\end{tcolorbox}
\caption{Pseudocode for the momentum-informed $\mu$TRec algorithm. The no-momentum version is detailed in Ref.~\cite{ughade_trec_2025}.}
\label{fig:mutrec-pseudocode}
\end{figure}

In this work, we focus on the incorporation of momentum information into the $\mu$TRec framework. Under the small angle approximation, MCS induces correlated fluctuations in $(y,\theta_y)$ that are well approximated by a zero mean bivariate Gaussian distribution with covariance matrix,
\begin{equation}
\Sigma(z_0,z_1)=
\begin{bmatrix}
\sigma_y^2(z_0,z_1) & \sigma_{y\theta}(z_0,z_1) \\
\sigma_{y\theta}(z_0,z_1) & \sigma_\theta^2(z_0,z_1)
\end{bmatrix}.
\end{equation}
This representation captures both angular spread and the correlated lateral displacement accumulated between detector planes.

Up to the standard MCS prefactor, the covariance elements over a segment $(z_0,z_1)$ are
\begin{align}
\sigma_y^2(z_0,z_1) &\propto \int_{z_0}^{z_1}\frac{(z_1-z)^2}{\beta^2(z)\,p^2(z)}\frac{dz}{X_0(z)}, \label{eq:sigy_def}\\
\sigma_{y\theta}(z_0,z_1) &\propto \int_{z_0}^{z_1}\frac{(z_1-z)}{\beta^2(z)\,p^2(z)}\frac{dz}{X_0(z)}, \label{eq:sigytheta_def}\\
\sigma_\theta^2(z_0,z_1) &\propto \int_{z_0}^{z_1}\frac{1}{\beta^2(z)\,p^2(z)}\frac{dz}{X_0(z)}. \label{eq:sigtheta_def}
\end{align}

Assuming a constant radiation length $X_0$ within each segment, momentum loss is introduced through a linearized Bethe--Bloch based model \cite{ughade_trec_2025},
\begin{equation}
p(z)\approx p_0-a z,
\end{equation}
where $p_0$ is the muon momentum at the segment entrance and $a$ is the average momentum loss per unit path length. For relativistic muons, $\beta(z)\approx 1$, and Eq.~\eqref{eq:sigtheta_def} becomes
\begin{align}
\sigma_\theta^2(z_0,z_1) &\propto \int_{z_0}^{z_1}\frac{dz}{\left(p_0-a z\right)^2} \nonumber\\
&= \frac{1}{p_0}\left[\frac{z_1}{p_0-a z_1}-\frac{z_0}{p_0-a z_0}\right] \nonumber\\
&= \frac{z_1-z_0}{p_0^2-a p_0 (z_1-z_0)+a^2 z_0 z_1}.
\label{eq:sigtheta_energyloss}
\end{align}
Under the weak loss condition $a(z_1-z_0)\ll p_0$, this reduces to
\begin{equation}
\sigma_\theta^2(z_0,z_1)\propto \frac{z_1-z_0}{p_0^2-a p_0 (z_1-z_0)}.
\end{equation}

Similarly, the remaining moments are expressed as \cite{Schultz_2007}
\begin{equation}
\sigma_y^2(z_0,z_1)\approx \frac{(z_1-z_0)^2}{3}\,\sigma_\theta^2(z_0,z_1),
\end{equation}
\begin{equation}
\sigma_{y\theta}(z_0,z_1)\approx \frac{\sqrt{3}}{2}\sqrt{\sigma_\theta^2(z_0,z_1)\,\sigma_y^2(z_0,z_1)}.
\end{equation}
These relations are exact for constant scattering power and remain accurate when the scattering power varies slowly over the segment, which is satisfied for the small step sizes used in the present reconstruction.

\subsection*{M-value}

For image formation, each voxel accumulates a scattering metric from the muons whose reconstructed trajectories intersect that voxel. Specifically, for a given voxel, the scattering angles from all muons passing through it are aggregated and the voxel intensity is taken as the mean scattering angle over the corresponding event set \cite{ughade_trec_2025}. In this study, we extend the conventional scattering-angle mapping by incorporating muon momentum information. This enables construction of density maps using the event-wise M-value, which combines the measured scattering angle and momentum and is compared directly against angle-only reconstructions. The M-value is defined as \cite{bae_momentum_2024, bae_fieldable_2022, bae_image_2024, bae_momentum-dependent_2022},

\begin{equation}\label{Mvalue}
M(p,\theta)=\log_{10}\!\left(\left(\theta^{0.958}\,[\mathrm{rad}] \times p\,[\mathrm{GeV}/c]\right)^{2.4}\right)
\end{equation}

Where, $\theta$ is the muon scattering angle and $p$ is the muon momentum.

\subsection*{Detector resolution}

In a typical tracking detector, the hit depth ($z$) is fixed by the detector plane location, whereas the transverse coordinates ($x$ and $y$) are limited by the readout granularity. Accordingly, we model the dominant measurement uncertainty by smearing only the transverse hit coordinates. Two representative spatial resolutions are considered: $5~\mathrm{mm}$ and $10~\mathrm{mm}$ full width at half maximum (FWHM). For each hit, independent Gaussian offsets are added to $x$ and $y$, with the Gaussian standard deviation given by
\[
\sigma=\frac{\mathrm{FWHM}}{2\sqrt{2\ln 2}}=\frac{\mathrm{FWHM}}{2.355}.
\]
This corresponds to $\sigma=2.12~\mathrm{mm}$ for $\mathrm{FWHM}=5~\mathrm{mm}$ and $\sigma=4.25~\mathrm{mm}$ for $\mathrm{FWHM}=10~\mathrm{mm}$.

For muon momentum (energy) measurement, we similarly examine two resolution cases, $5\%$ and $10\%$, quoted as relative FWHM. For a muon with energy $E_\mu$, the assumed $\mathrm{FWHM}=r\,E_\mu$, where $r\in\{0.05,\,0.10\}$, and the corresponding Gaussian width is
\[
\sigma_E=\frac{\mathrm{FWHM}}{2.355}=\frac{r\,E_\mu}{2.355}.
\]

\section*{Results and Discussion}

Multiple test scenarios were tested to evaluate the $\mu$TRec algorithm under operating conditions relevant to practical deployment. An overview of the scenarios and associated test conditions is provided in Table \ref{tab:scenarios}. For momentum-informed case, each voxel value is computed as the mean of the $M$-values associated with all muons traversing that voxel. Similarly, for no-momentum information case, each voxel value is computed as the mean of the scattering-angle values associated with all muons traversing that voxel. For core-level analysis, a 2D representation is obtained by averaging the reconstructed 3D voxel matrix along the microreactor axial direction. For quantitative evaluation, we focus on identifying a single missing hexagonal fuel flake in a reactor core configuration consisting of 61 flakes. The resulting separation between the intact and missing-flake responses is quantified using the detectability metric based on the difference of means,

\begin{equation}
\mathrm{DP}_{\mathrm{means}} =
\frac{\left|\mu_{\mathrm{intact}} - \mu_{\mathrm{miss}}\right|}
{\sqrt{\sigma_{\mathrm{intact}}^{2}/n_{\mathrm{intact}} + \sigma_{\mathrm{miss}}^{2}/n_{\mathrm{miss}}}}
\end{equation}

where $\mathrm{DP}_{\mathrm{means}}$ is the detectability metric based on the separation of the two sample means, $\mu_{\mathrm{intact}}$ and $\mu_{\mathrm{miss}}$ are the mean voxel values in the intact-fuel and missing-fuel regions, respectively, $\sigma_{\mathrm{intact}}$ and $\sigma_{\mathrm{miss}}$ are the corresponding standard deviations, and $n_{\mathrm{intact}}$ and $n_{\mathrm{miss}}$ are the numbers of voxels used to compute the statistics in each region. 

$\mu$TRec is a physics-informed reconstruction algorithm that, in its standard form, relies on three key assumptions: (1) the radiation length of the materials traversed by the muon, (2) the momentum of each muon, and (3) the energy loss of each muon along its path. In this study, we relax assumptions (2) and (3) by incorporating two spectrometers and directly measuring the muon energy at the entry and exit of the system. The upstream spectrometer provides the incident muon energy $E_1$, while the downstream spectrometer provides the exiting energy $E_2$, enabling the energy loss to be estimated as $E_1 - E_2$. With these measurements, the only remaining model input is the radiation length of the intervening materials. However, the use of two spectrometers increases system cost and integration complexity. We therefore also evaluate a reduced-instrumentation configuration that uses a single spectrometer, in which case the muon energy loss must be treated as an assumed or modeled quantity rather than a directly measured one.

\begin{table}[ht]
\centering
\renewcommand{\arraystretch}{1.15}
\setlength{\tabcolsep}{6pt}
\begin{tabular}{|c|l|p{0.43\linewidth}|}
\hline
\textbf{No.} & \textbf{Scenario} & \textbf{Configuration} \\
\hline
1 & Laser-based muon source & Momentum-informed vs.\ without-momentum reconstruction across four voxel sizes for 3M muons \\
\hline
2 & Cosmic-ray muon source & Momentum-informed vs.\ without-momentum reconstruction across four voxel sizes for 3M muons \\
\hline
3 & $\mu$TRec vs.\ PoCA comparison & Momentum-informed vs.\ without-momentum reconstruction across four muon counts at 10 mm voxel size \\
\hline
4 & Detector resolution (momentum-informed) & Cosmic-ray muons at 4\,mm voxel size for 3M muons\\
\hline
5 & Number of spectrometers & Cosmic-ray muons at 4\,mm voxel size for 3M muons \\
\hline
\end{tabular}
\caption{\label{tab:scenarios}Test scenarios evaluated in this study.}
\end{table}

\subsection*{Laser-based muon source}

\begin{figure*}[h]
    \centering
      \label{fig:a}%
      \includegraphics[width=0.25\textwidth]{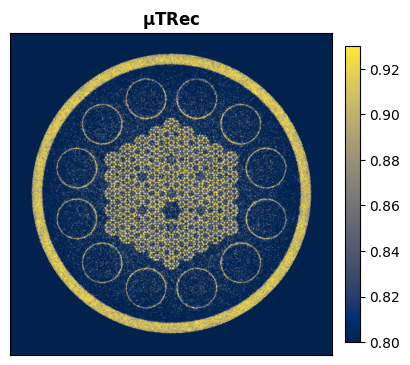}%
    \hfill
      \label{fig:b}%
      \includegraphics[width=0.25\textwidth]{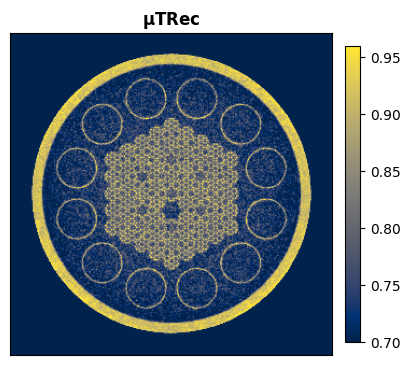}%
    \hfill
      \label{fig:c}%
      \includegraphics[width=0.25\textwidth]{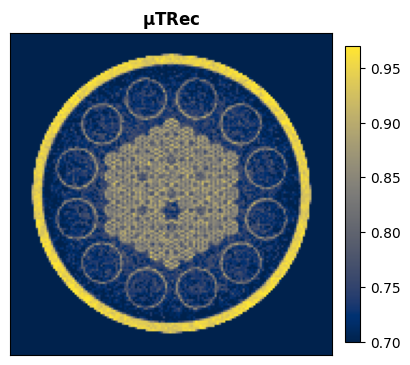}%
    \hfill
      \label{fig:d}%
      \includegraphics[width=0.25\textwidth]{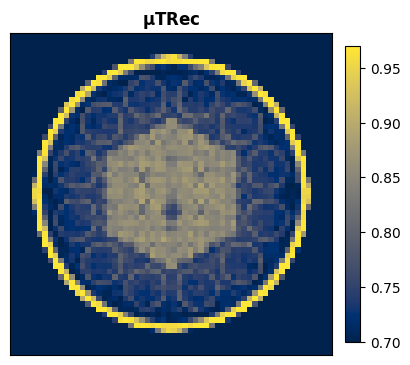}%

    \subfloat[Voxel size: 4 mm]{%
      \label{fig:e}%
      \includegraphics[width=0.25\textwidth]{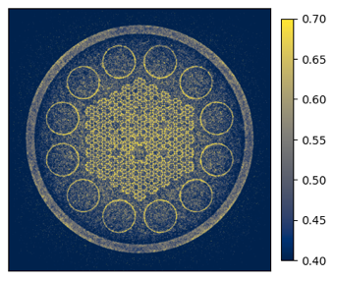}%
    }%
    \hfill
    \subfloat[Voxel size: 10 mm]{%
      \label{fig:f}%
      \includegraphics[width=0.25\textwidth]{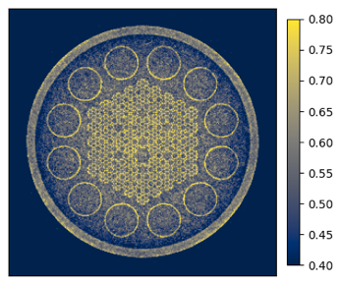}%
    }%
    \hfill
    \subfloat[Voxel size: 20 mm]{%
      \label{fig:g}%
      \includegraphics[width=0.25\textwidth]{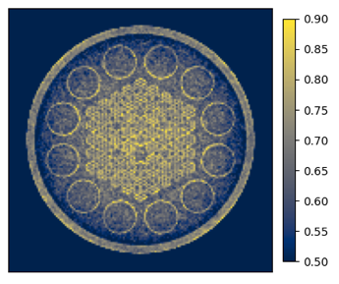}%
    }%
    \hfill
    \subfloat[Voxel size: 50 mm]{%
      \label{fig:h}%
      \includegraphics[width=0.25\textwidth]{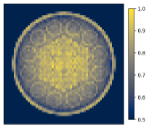}%
    }%
    
    \caption{\textbf{Laser muon source:} reconstructions using the $\mu$TRec algorithm. Top row shows reconstructions using entry and exit momentum-information with M-value mapping, bottom row shows reconstructions without momentum using scattering-angle-based density mapping.}
    \label{results:laser}
\end{figure*}

To evaluate the $\mu$TRec algorithm, the first test scenario assumes an idealized laser-driven muon source, producing a parallel beam of 5 GeV muons. Laser-based muon sources are an active area of research in high-energy-density physics, with several organizations investigating their potential for muon tomography applications \cite{Terzani_2025, Calvin_2023, Ludwig_2025}. The resulting microreactor reconstructions for core evaluation with 3M muons are shown in Fig.~\ref{results:laser}. The top row presents momentum-informed reconstructions, and the bottom row presents reconstructions performed without momentum information, for voxel sizes of (a) 4 mm, (b) 10 mm, (c) 20 mm, and (d) 50 mm. As expected, image noise increases with voxel size for both configurations, and the 4 mm voxel grid yields the best image quality. In the momentum-informed case, the missing flake is readily identifiable at all four voxel sizes. In addition, key core features are resolved, including the seven shutdown rods (hollow, 0.3 cm stainless-steel wall thickness) arranged in a hexagonal pattern and the 12 circular control drums (hollow, 0.65 cm lead wall thickness). The smaller blue circular features correspond to heat pipes (hollow, 1 mm stainless-steel wall thickness). The UO$_2$ fuel pins, with a diameter of 1.7 cm, are consistently reconstructed across all 61 flakes and appear as the yellow regions in the core. In the absence of momentum information, key features such as the shutdown rods, fuel pins, and heat pipes become difficult to resolve at a 20 mm voxel size, and the reconstruction at a 50 mm voxel size is no longer reliable for identifying a missing fuel flake. These trends are quantified in Table~\ref{tab:image_characteristics}. Compared with the momentum-missing configuration, the momentum-informed reconstruction increases the DP for detecting a missing fuel flake by 144.22\%, 141.03\%, 149.85\%, and 65.50\% for voxel sizes of 4~mm, 10~mm, 20~mm, and 50~mm, respectively.

\begin{table}[ht]
\centering
\begin{tabular}{|l|l|c|c|c|c|}
\hline
\textbf{Source} & \textbf{Setting} & \textbf{4mm} & \textbf{10mm} & \textbf{20mm} & \textbf{50mm} \\
\hline
\multirow{2}{*}{Laser muon source} 
  & With momentum    & 51.8318 & 28.1355 & 13.0012 & 7.4380 \\
\cline{2-6}
  & Without momentum & 21.2233 & 11.6730 & 5.2037  & 4.4943 \\
\cline{2-6}
  & \textbf{Improvement in DP} & \textbf{144.22$\%$} & \textbf{141.03$\%$} & \textbf{149.85$\%$}  & \textbf{65.50$\%$} \\
\hline
\multirow{2}{*}{Cosmic muon source} 
  & With momentum    & 59.1885 & 33.5523 & 16.9713 & 5.8577 \\
\cline{2-6}
  & Without momentum & 28.8570 & 18.1625 & 11.4346 & 4.9245 \\
\cline{2-6}
  & \textbf{Improvement in DP} & \textbf{105.11$\%$} & \textbf{84.73$\%$} & \textbf{48.42$\%$}  & \textbf{18.95$\%$} \\
\hline
\end{tabular}
\caption{\label{tab:image_characteristics}Reconstructed image characteristics (detectability metric, DP) using a laser muon source and a realistic energy spectrum with 3 $\times$ $10^{6}$ muons for a single missing fuel assembly using $\mu$TRec algorithm, reported for different voxel sizes.}
\end{table}

\subsection*{Cosmic-ray muon source}

\begin{figure*}[h]
    \centering
      \label{fig:a}%
      \includegraphics[width=0.25\textwidth]{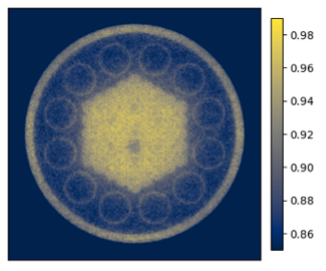}%
    \hfill
      \label{fig:b}%
      \includegraphics[width=0.25\textwidth]{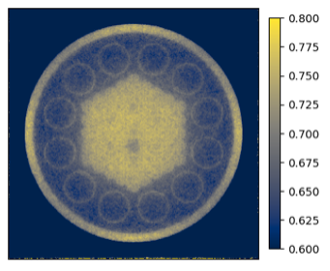}%
    \hfill
      \label{fig:c}%
      \includegraphics[width=0.25\textwidth]{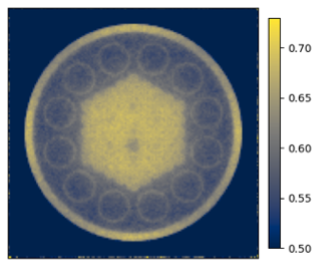}%
    \hfill
      \label{fig:d}%
      \includegraphics[width=0.25\textwidth]{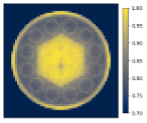}%

    
    \subfloat[Voxel size: 4 mm]{%
      \label{fig:e}%
      \includegraphics[width=0.25\textwidth]{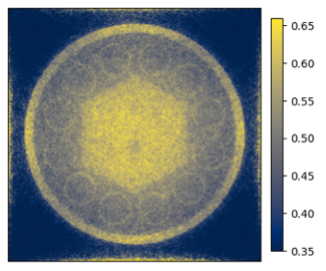}%
    }%
    \hfill
    \subfloat[Voxel size: 10 mm]{%
      \label{fig:f}%
      \includegraphics[width=0.25\textwidth]{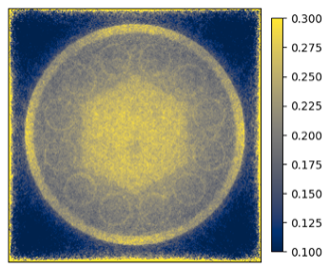}%
    }%
    \hfill
    \subfloat[Voxel size: 20 mm]{%
      \label{fig:g}%
      \includegraphics[width=0.25\textwidth]{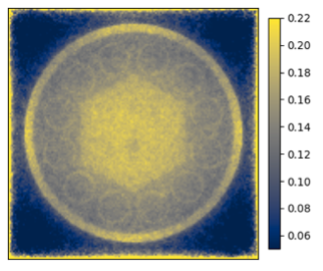}%
    }%
    \hfill
    \subfloat[Voxel size: 50 mm]{%
      \label{fig:h}%
      \includegraphics[width=0.25\textwidth]{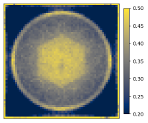}%
    }%
    
    \caption{\textbf{Cosmic-ray muon source:} reconstructions using the $\mu$TRec algorithm. Top row shows reconstructions using entry and exit momentum-information with M-value mapping, bottom row shows reconstructions without momentum using scattering-angle-based density mapping.}
    \label{results:cosmic}
\end{figure*}

In the second scenario, we consider a cosmic-ray muon source, with the corresponding reconstructions for 3M muons shown in Fig.~\ref{results:cosmic}. As in the previous scenario, the top row presents momentum-informed results and the bottom row presents momentum-missing results for the same four voxel sizes. As expected, increasing the voxel size leads to higher noise levels in both configurations. Overall, the momentum-informed reconstructions provide improved image quality and more reliable identification of a missing fuel flake relative to the momentum-missing case. For the cosmic-ray source, distinguishing fuel pins from heat pipes remains challenging even when momentum information is included. Despite this limitation, the momentum-informed reconstructions allow identification of a missing flake across all voxel sizes. In contrast, the momentum-missing reconstructions only weakly indicate the missing flake, and at a 50~mm voxel size the feature cannot be identified with confidence. Control drums are discernible in both configurations. Shutdown rods are only weakly resolved in the momentum-informed case and are not identifiable in the momentum-missing case. Quantitatively, Table~\ref{tab:image_characteristics} shows that momentum-informed reconstruction increases DP by 105.11\%, 84.73\%, 48.42\%, and 18.95\% for voxel sizes of 4~mm, 10~mm, 20~mm, and 50~mm, respectively, compared with the momentum-missing configuration.

\begin{figure}[h]
\centering
\begin{subfigure}[b]{0.47\linewidth}
    \centering
    \includegraphics[width=\linewidth]{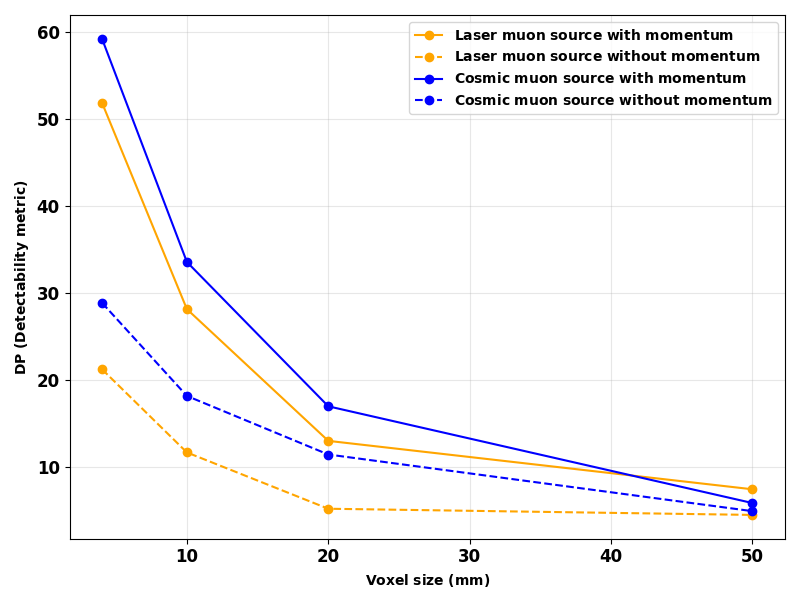}
    \caption{}
    \label{change_in_DP_a}
\end{subfigure}
\hfill
\begin{subfigure}[b]{0.47\linewidth}
    \centering
    \includegraphics[width=\linewidth]{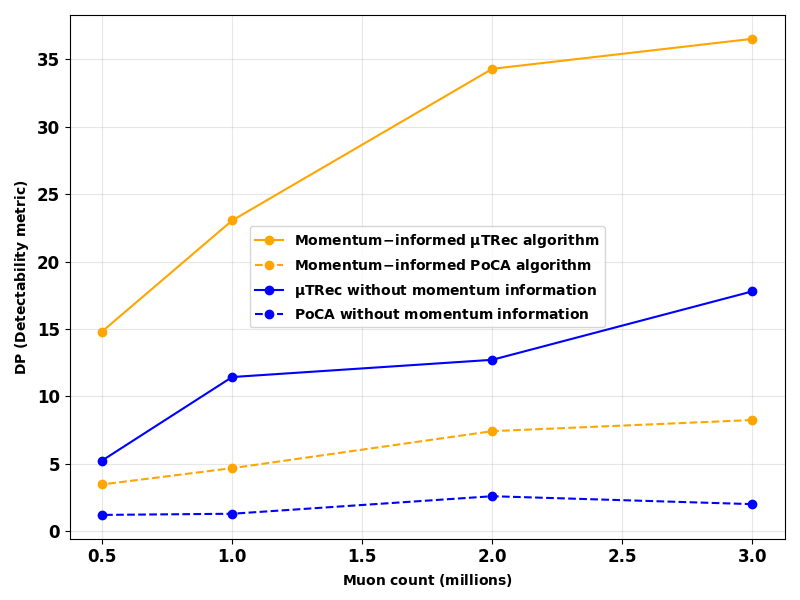}
    \caption{}
    \label{change_in_DP_b}
\end{subfigure}

\caption{Detection performance for different test scenarios: \textbf{(a) DP vs. voxel-size} for 3M muons in the single missing fuel flake scenario using $\mu$TRec algorithm and \textbf{(b) DP vs. muon count}, comparing $\mu$TRec and PoCA for momentum-informed and no-momentum cosmic ray cases with 10 mm voxels. }
\label{fig:change_in_DP}
\end{figure}

Figure \ref{change_in_DP_a} shows that DP decreases as voxel size increases. For both the laser-driven and cosmic muon sources, the separation between the momentum-informed and no-momentum cases is largest at small voxel sizes and narrows as voxel size increases. This indicates that event-level momentum information provides the greatest benefit when high spatial resolution is required. At finer voxelization, accurate assignment of each muon trajectory to the correct set of voxels becomes more sensitive to trajectory uncertainty. Incorporating momentum reduces this uncertainty in the scattering model, leading to more accurate trajectory estimates and improved detectability. 
We observe higher DP for cosmic ray muons than for the laser muon source in both the momentum-informed and no-momentum cases. This is because the laser driven reconstructions exhibit stronger voxel level fluctuations within the intact fuel and missing fuel regions, which increases the corresponding standard deviations and reduces DP.

\subsection*{$\bm{\mu}$TRec vs.\ PoCA comparison}

\begin{figure*}[t]
    \centering
      \label{fig:a}%
      \includegraphics[width=0.25\textwidth]{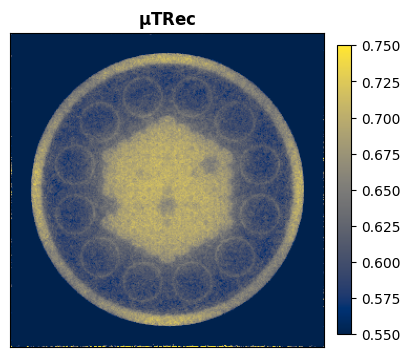}%
    \hfill
      \label{fig:b}%
      \includegraphics[width=0.25\textwidth]{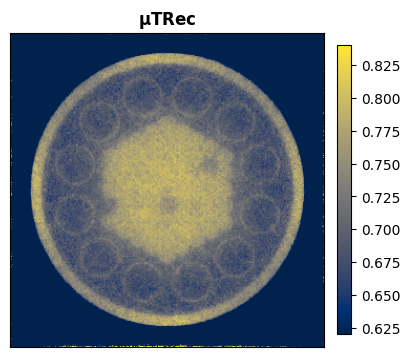}%
    \hfill
      \label{fig:c}%
      \includegraphics[width=0.25\textwidth]{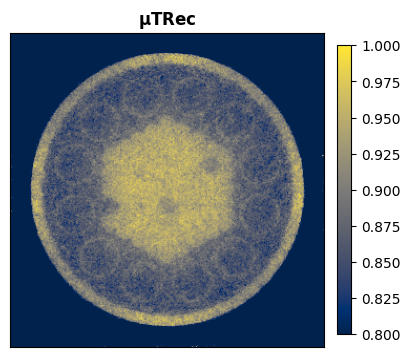}%
    \hfill
      \label{fig:d}%
      \includegraphics[width=0.25\textwidth]{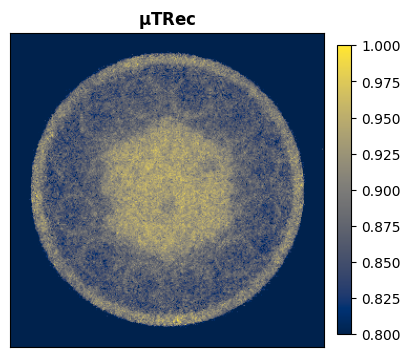}%

    
    \subfloat[3M muons]{%
      \label{fig:e}%
      \includegraphics[width=0.25\textwidth]{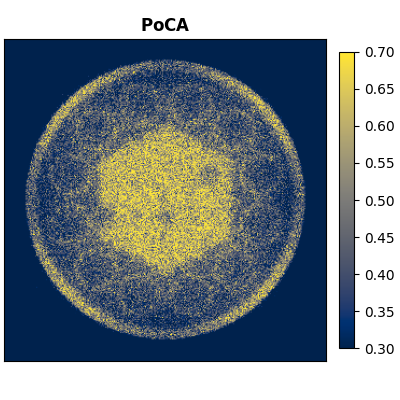}%
    }%
    \hfill
    \subfloat[2M muons]{%
      \label{fig:f}%
      \includegraphics[width=0.25\textwidth]{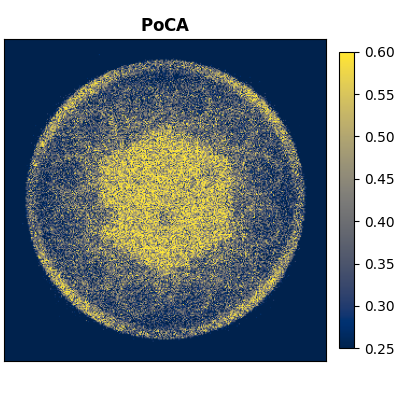}%
    }%
    \hfill
    \subfloat[1M muons]{%
      \label{fig:g}%
      \includegraphics[width=0.25\textwidth]{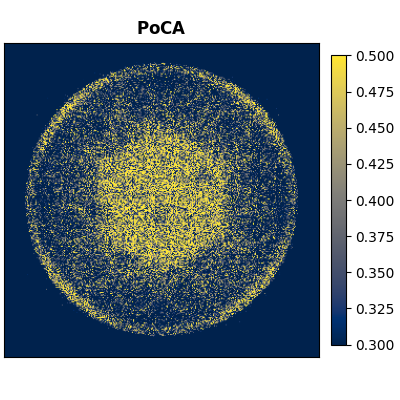}%
    }%
    \hfill
    \subfloat[0.5M muons]{%
      \label{fig:h}%
      \includegraphics[width=0.25\textwidth]{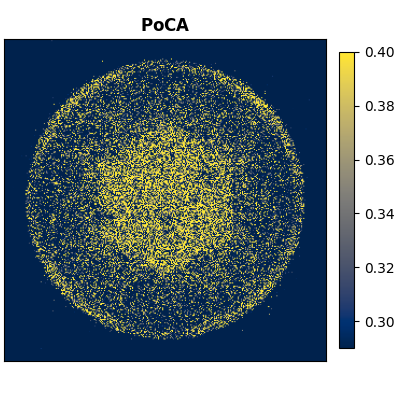}%
    }%
    
    \caption{\textbf{Momentum-informed:} Cosmic-ray muon source, top row shows reconstructions using $\mu$TRec with M-value mapping, bottom row shows reconstructions with PoCA using M-value mapping, for 10 mm voxels.}
    \label{results:poca_comparison_momentum}
\end{figure*}

\begin{figure*}[t]
    \centering
      \label{fig:a}%
      \includegraphics[width=0.25\textwidth]{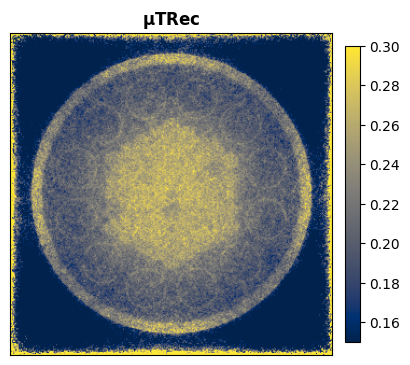}%
    \hfill
      \label{fig:b}%
      \includegraphics[width=0.25\textwidth]{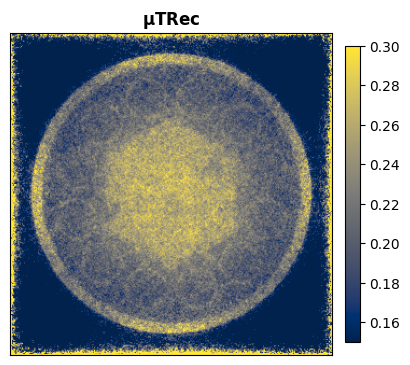}%
    \hfill
      \label{fig:c}%
      \includegraphics[width=0.25\textwidth]{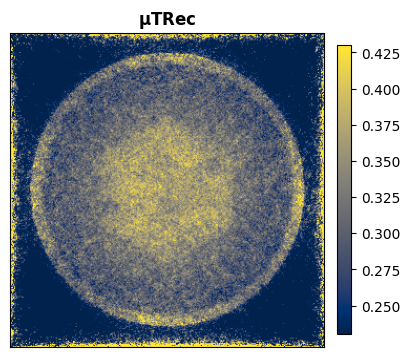}%
    \hfill
      \label{fig:d}%
      \includegraphics[width=0.25\textwidth]{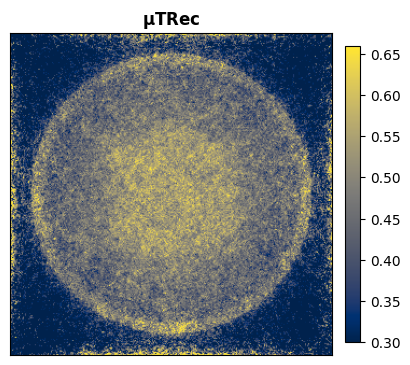}%

    
    \subfloat[3M muons]{%
      \label{fig:e}%
      \includegraphics[width=0.25\textwidth]{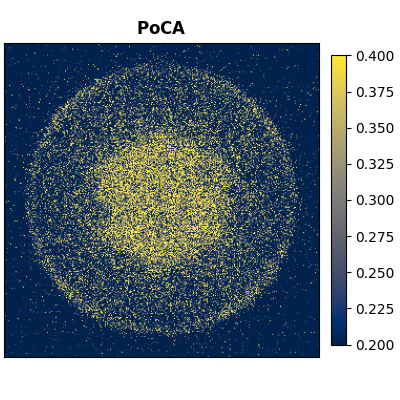}%
    }%
    \hfill
    \subfloat[2M muons]{%
      \label{fig:f}%
      \includegraphics[width=0.25\textwidth]{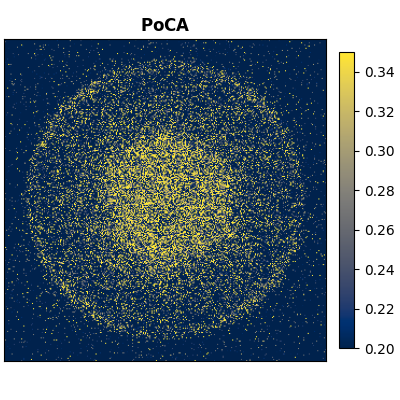}%
    }%
    \hfill
    \subfloat[1M muons]{%
      \label{fig:g}%
      \includegraphics[width=0.25\textwidth]{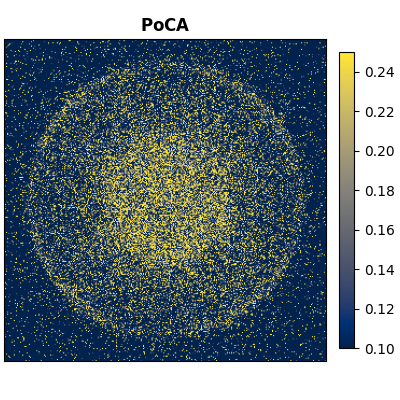}%
    }%
    \hfill
    \subfloat[0.5M muons]{%
      \label{fig:h}%
      \includegraphics[width=0.25\textwidth]{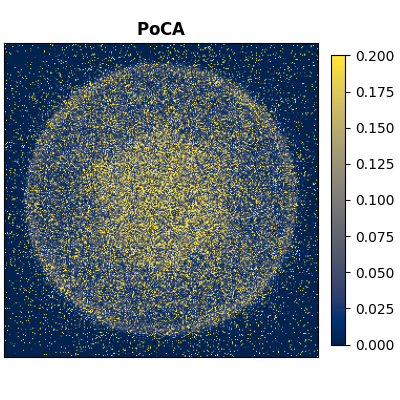}%
    }%
    
    \caption{\textbf{No-momentum information:} Cosmic-ray muon source, Top row shows reconstructions using $\mu$TRec with scattering angle mapping, bottom row shows reconstructions with PoCA using scattering angle mapping, for 10 mm voxels.}
    \label{results:poca_comparison_no_momentum}
\end{figure*}

In the third scenario, we benchmark $\mu$TRec against the widely used PoCA algorithm for cosmic ray muons, considering both the no-momentum case and the momentum informed case. We evaluate four event totals, 3, 2, 1, and 0.5 million muons. In cosmic ray operation, larger muon counts correspond to longer acquisition times, so the goal is to identify microreactor defects with as few muons as possible to reduce total imaging time. For both reconstruction methods, runtimes are on the order of a few seconds when implemented in JAX with GPU acceleration. In the momentum informed comparison, both algorithms use measured momentum for $M$-value mapping, while $\mu$TRec additionally incorporates per-muon momentum and energy loss within the trajectory estimation. Figure~\ref{results:poca_comparison_momentum} presents momentum-informed reconstructions at a fixed 10 mm voxel size, with $\mu$TRec shown in the top row and PoCA in the bottom row. The defect model includes three missing fuel flakes, including one near the core edge. For $\mu$TRec, the 3M, 2M, and 1M cases clearly resolve all three missing flakes and the control drums. At 0.5M muons, the edge flake becomes difficult to identify and the control drums begin to blur, while the other two missing flakes remain discernible. PoCA detects these three missing flakes at 3M and 2M muons, but the reconstructions remain noisy and the control drums are poorly resolved. At 1M and 0.5M muons, the missing flakes cannot be identified with confidence, and the control drums are not distinguishable. The no-momentum case is more challenging, as shown in Figure~\ref{results:poca_comparison_no_momentum}. With $\mu$TRec, all three missing flakes are most apparent in the 3M case. At 2M and 1M muons, the indications are weaker but still potentially detectable, while the 0.5M case does not support reliable identification of any flake or the control drums. For PoCA, all four muon counts in the no-momentum setting, fail to identify missing flakes or resolve the control drums.


Quantitatively, Table~\ref{tab:poca} shows that, in the momentum informed case, $\mu$TRec increases DP by 342.54\%, 361.75\%, 392.14\%, and 326.13\% relative to PoCA for 3M, 2M, 1M, and 0.5M muons, respectively. For the no-momentum case, the corresponding DP improvements are 783.57\%, 388.72\%, 778.81\%, and 331.59\%. Figure~\ref{change_in_DP_b} further summarizes this trend, showing that $\mu$TRec consistently provides higher detectability than PoCA, with the advantage becoming more pronounced at higher muon counts.

\begin{table}[ht]
\centering
\begin{tabular}{|l|c|c|c|c|c|c|}
\hline
\multirow{2}{*}{\textbf{\shortstack{Muon fluence\\(millions)}}} &
\multicolumn{2}{c|}{\textbf{DP (No-momentum)}} & \textbf{Improvement in DP} &
\multicolumn{2}{c|}{\textbf{DP (Momentum-informed)}} &
\textbf{Improvement in DP} \\
\cline{2-3} \cline{5-6}
 & \textbf{PoCA} & \textbf{$\bm{\mu}$TRec} &
\textbf{($\bm{\mu}$TRec vs. PoCA)} & \textbf{PoCA} & \textbf{$\bm{\mu}$TRec} & \textbf{($\bm{\mu}$TRec vs. PoCA)} \\
\hline
3M muons   & 2.0134 & 17.7897 & \textbf{783.57\%} & 8.2498 & 36.5089 & \textbf{342.54\%} \\
\hline
2M muons   & 2.6013 & 12.7131 & \textbf{388.72\%} & 7.4247 & 34.2834 & \textbf{361.75\%} \\
\hline
1M muons   & 1.3013 & 11.4359 & \textbf{778.81\%} & 4.6816 & 23.0402 & \textbf{392.14\%} \\
\hline
0.5M muons & 1.2149 & 5.2434 & \textbf{331.59\%} & 3.4748 & 14.8071 & \textbf{326.13\%} \\
\hline
\end{tabular}
\caption{DP comparison between PoCA and $\mu$TRec using cosmic-ray muons for no-momentum and momentum-informed cases with 10 mm voxels.}
\label{tab:poca}
\end{table}

\subsection*{Detector resolution}

\begin{figure}[ht]
\centering
\begin{subfigure}[b]{0.32\linewidth}
    \centering
    \includegraphics[width=\linewidth]{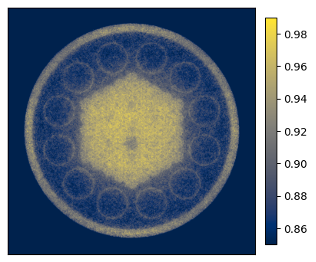}
    \caption{Ideal resolution}
    \label{fig:a}
\end{subfigure}
\hfill
\begin{subfigure}[b]{0.32\linewidth}
    \centering
    \includegraphics[width=\linewidth]{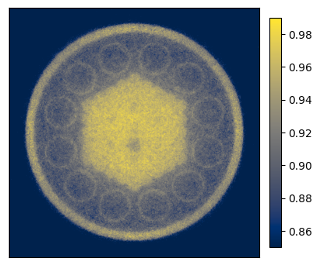}
    \caption{5 mm spatial and 5$\%$ energy resolution }
    \label{fig:b}
\end{subfigure}
\hfill
\begin{subfigure}[b]{0.32\linewidth}
    \centering
    \includegraphics[width=\linewidth]{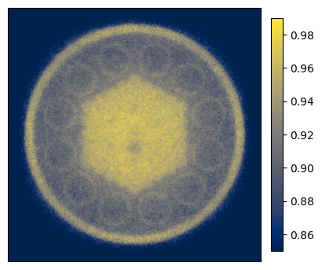}
    \caption{10 mm spatial and 10$\%$ energy resolution}
    \label{fig:c}
\end{subfigure}

\caption{Effect of detector resolution on image quality.}
\label{fig:resolution}
\end{figure}

The impact of detector resolution is a key consideration for realistic deployment. Here, we assess the sensitivity of $\mu$TRec to finite spatial (position) resolution and energy resolution using 3M cosmic ray muons for a single missing fuel flake case. As shown in Fig.~\ref{fig:resolution}, three detector configurations are considered: (a) ideal resolution, (b) 5~mm spatial resolution with 5\% energy resolution, and (c) 10~mm spatial resolution with 10\% energy resolution. As expected, reconstruction noise increases from case (a) to case (c). Quantitatively, Table~\ref{tab:resolution_dp} shows that, relative to the ideal-resolution case, DP decreases by 0.7\% and 8.88\% for cases (b) and (c), respectively. Nevertheless, the missing fuel flake and the control drums remain detectable in all three cases. These results indicate that $\mu$TRec is robust to realistic detector position and energy resolution.

\begin{table}[ht]
\centering
\begin{tabular}{|l|c|c|}
\hline
 \textbf{Detector resolution} & \textbf{DP} & \textbf{Reduction in DP} \\
\hline
Ideal resolution & 59.1885 & - \\
\hline
5mm spatial and $5\%$ energy resolution & 58.7726 & 0.7$\%$\\
\hline
10mm spatial and $10\%$ energy resolution & 53.9325 & 8.88$\%$\\
\hline
\end{tabular}
\caption{\label{tab:resolution_dp} DP for different detector resolution configurations at 4~mm voxel size using 3M cosmic-ray muons.}
\end{table}

\subsection*{Number of spectrometers}

Table~\ref{tab:spectrometer_dp} summarizes three $\mu$TRec configurations. In the first configuration, muon momentum is not available and the density map is constructed using only the scattering angle. In the second configuration, a single upstream spectrometer measures the incident muon momentum, which is then used as an input to $\mu$TRec and for $M$-value based density mapping. In the third configuration, two spectrometers measure the incident and exiting muon momenta, enabling an event-by-event estimate of energy loss that is incorporated into $\mu$TRec in addition to the incident momentum. Relative to the no-spectrometer configuration, the one-spectrometer configuration increases DP by 101.06\%. Using two spectrometers increases DP by 105.11\%. The additional gain from a second spectrometer is therefore modest compared with the one-spectrometer case. Given the added cost and integration complexity of a second spectrometer, these results suggest that a single spectrometer provides an effective performance--cost trade-off for practical implementation.

\begin{table}[ht]
\centering
\begin{tabular}{|l|c|c|}
\hline
 \textbf{Configuration} & \textbf{DP} & \textbf{Improvement in DP} \\
\hline
No spectrometer & 28.8570 & - \\
\hline
1 spectrometer & 58.0199 & 101.06$\%$\\
\hline
2 spectrometers & 59.1885 & 105.11$\%$\\
\hline
\end{tabular}
\caption{\label{tab:spectrometer_dp}DP for different spectrometer configurations at 4~mm voxel size using 3M cosmic-ray muons.}
\end{table}


\section*{Conclusion}

In this work, we presented and validated $\mu$TRec, a physics-informed reconstruction algorithm for muon scattering tomography that explicitly models multiple Coulomb scattering and incorporates per-muon kinematics through a Bayesian trajectory estimation framework. The central contribution of this study is a systematic evaluation of momentum-informed $\mu$TRec for microreactor core imaging, with emphasis on how momentum measurements improve defect detectability and feature fidelity under conditions relevant to practical deployment.

Across both an idealized laser-driven source and a cosmic-ray source, momentum-informed $\mu$TRec consistently produced higher quality reconstructions and improved missing-flake detectability relative to reconstructions performed without momentum information. For the laser-driven source, incorporating momentum information increased DP by 144.22\%, 141.03\%, 149.85\%, and 65.50\% for voxel sizes of 4 mm, 10 mm, 20 mm, and 50 mm, respectively. For the cosmic-ray source, the corresponding DP improvements were 105.11\%, 84.73\%, 48.42\%, and 18.95\%. These results show that momentum-informed $\mu$TRec provides a clear and repeatable advantage, especially at finer voxelization, where event-level scattering statistics are most important for preserving small-scale structural signatures. We also compared $\mu$TRec with PoCA under realistic cosmic-ray conditions at 10 mm voxel resolution over multiple muon counts. In the momentum-informed case, $\mu$TRec increased DP by 342.54\%, 361.75\%, 392.14\%, and 326.13\% relative to PoCA for 3M, 2M, 1M, and 0.5M muons, respectively. In the no-momentum case, the corresponding improvements were 783.57\%, 388.72\%, 778.81\%, and 331.59\%. These comparisons confirm that the physics-informed trajectory model in $\mu$TRec yields substantially better detectability than point-based reconstruction, even when muon statistics are limited.

We further examined practical instrumentation choices for momentum-informed operation. Adding a single upstream spectrometer increased DP by 101.06\% relative to the no-spectrometer configuration, while adding a second spectrometer increased DP by 105.11\%. The modest incremental benefit from the second spectrometer suggests that a single-spectrometer configuration can capture most of the performance gain while reducing system cost and integration complexity. In addition, momentum-informed $\mu$TRec remained robust to realistic detector resolution. Relative to the ideal-resolution case, DP decreased by only 0.7\% and 8.88\% for detector configurations with 5 mm and 10 mm spatial resolution paired with 5\% and 10\% energy resolution, respectively. Overall, these results support momentum-informed $\mu$TRec as a practical and scalable approach for non-intrusive verification of sealed microreactor cores.

\section*{Data availability}

Python and MATLAB scripts for $\mu$TRec and PoCA algorithms are publicly available at \url{https://github.com/rughade/muTRec-algorithm}. 

\section*{Acknowledgements}

The present research is being conducted with the support of funding provided by the Purdue University School of Nuclear Engineering and the Purdue Research Foundation. 

\bibliography{sample.bib}

\end{document}